\documentclass[aps,12pt,a4paper]{revtex4-1}
\usepackage{textcomp}
\usepackage{graphicx}
\usepackage{amsmath}
\usepackage{wrapfig}
\usepackage[usenames,dvipsnames]{color}

\begin{document}

\title{Defect-induced shift of the Peierls transition in TTF-TCNQ thin films}
\author{Vita Solovyeva and Michael Huth}
\affiliation{Physikalisches Institut, Goethe Universit\"{a}t,\\Max-von-Laue-Str. 1, 60438 Frankfurt am Main, Germany\\E-mail: levitan@physik.uni-frankfurt.de}

\begin{abstract}

In this paper we investigate the influence of the substrate material and film thickness  on the Peierls transition temperature in tetrathiafulvalene-tetracyanoquinodimethane (TTF-TCNQ) thin films, grown by physical vapor deposition. Our analysis shows that the substrate material and the growth conditions strongly influence the film morphology.
In particular, we demonstrate that the Peierls transition temperature  in thin films is lower than in TTF-TCNQ single crystals.
We argue that this effect arises due to defects, which emerge in TTF-TCNQ thin films during the growth process.

\end{abstract}


\maketitle

\section{Introduction}
\label{sec:introduction}

Organic charge transfer systems represent a material class for interdisciplinary research on the borderline of correlation physics, material science and chemistry \cite{Toyota07}. They also form the basis for extending the rapidly growing field of organic electronics towards binary donor-acceptor systems. In this regard thin film growth studies, as well as surface and interface oriented research on the electronic properties of these materials become more and more important \cite{Koch_devices,Mein_BEDT-TTF-TCNQ,Medjanik,Indranil,arxiv}. There are several aspects which can be studied in thin films that are not accessible in single crystals, such as interface- and surface-induced states, substrate-induced strain effects and the role of substrate-induced defects with regard to the electronic properties of these materials.

\begin{figure}[b]
\includegraphics[width=1\textwidth]{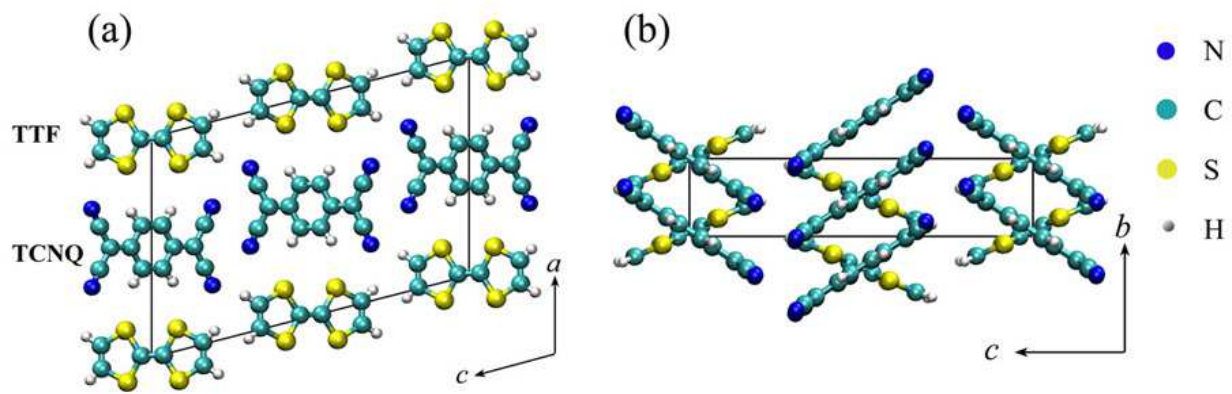}
\caption{Crystalline structure of TTF-TCNQ in the monoclinic crystal system with space group  P2$_1/$c. (a)  Orthogonal view along the stacking \textit{b}-axis. (b) Orthogonal view along the direction perpendicular to the \textit{(bc)}-plane of TTF-TCNQ.}
\label{fig:sturcture}
\end{figure}

The interest in the electronic properties of organic charge transfer materials was boosted already in 1973, after the first successful fabrication of a novel organic conductor, tetrathiaful\-valene-tetracyanoquinodimethane (TTF-TCNQ) \cite{TTF-TCNQ-first,TTF-TCNQ_giant}. TTF-TCNQ consists of parallel homosoric stacks of acceptor (TCNQ) and donor (TTF) molecules as  illustrated schematically in Fig.~\ref{fig:sturcture} \cite{cryst-str-TTF-TCNQ}. It was demonstrated that due to the interaction between the $\pi$-orbitals arising along the stack direction (corresponding to the {\it b-} direction in Fig.~\ref{fig:sturcture}b) the electrical conductivity of TTF-TCNQ is strongly anisotropic with $\sigma_b/\sigma_a>10^2$ at room temperature, where $\sigma_a$ and $\sigma_b$ are the electrical conductivities along the \textit{a}- and \textit{b}- directions, respectively \cite{conductivity_ratio}.
TTF-TCNQ single crystals show metallic behavior down to about 60~K and undergo a series of phase transitions at $T_H=54$~K, $T_I=49$~K and $T_L=38$~K \cite{Jerome_PRL_1978,Khanna,X_ray_scattering}, which successively suppress the metallic conductivity of the TTF and TCNQ chains, turning the material into an insulator. The phase transition at 54~K is driven by a charge density wave (CDW) Peierls instability in the TCNQ chains and is usually referred to as Peierls transition~\cite{Jerome_Schulz}.

In recent studies the hydrostatic pressure dependence of the TTF-TCNQ phase transitions was investigated \cite{Jerome_PRL_1978, Temperature-Pressure,Pressure_2009}. Pressure increases the overlap of the electronic wave functions between the TTF and TCNQ stacks and thus alleviates the quasi one-dimensional conductivity of TTF-TCNQ, which in turn suppresses the Peierls instability \cite{Jerome_2004} and presumably permits superconductivity~\cite{Temperature-Pressure}. Therefore, the dependence of the Peierls transition temperature of TTF-TCNQ on pressure is crucial for understanding the nature of this transition. It was demonstrated that at higher pressures ($\gtrsim$1.5~GPa) only one phase transition associated with the CDW exists \cite{Jerome_PRL_1978}. Later it was demonstrated that under extremely high pressures ($\gtrsim$ 3~GPa) a suppression of the CDWs is taking place \cite{Temperature-Pressure}.

The concentration of defects in the material is another important factor, which influences the CDWs in TTF-TCNQ. The increase of the concentration of defects in TTF-TCNQ single crystals leads to a lowering of the Peierls transition temperature, because the long-range order in the 1D~crystal becomes disturbed and the coherence length associated with CDW state is reduced \cite{1d-3d,Gruner,Chiang}. In \cite{Chiang} defects in a TTF-TCNQ single crystal were induced by 8~MeV deuteron irradiation of the sample. It was demonstrated that the Peierls transition temperature shifts towards lower temperature with increase of the radiation dose. The Peierls transition vanished when the defect concentration in the TTF-TCNQ single crystal reached a few percent \cite{Zuppiroli_irradiation}. In this case the electrical conductivity of the crystal showed thermo-activated behavior.

The factors influencing the properties of TTF-TCNQ single crystals were extensively investigated in \cite{Jerome_PRL_1978,Temperature-Pressure,Chu_pressure,Zuppiroli_irradiation,Chiang}. TTF-TCNQ thin films provide an excellent model system for investigation the influence of critical parameters on material properties, which cannot be studied on single crystals. For example, by applying uni- and biaxial strain to a TTF-TCNQ thin film one can tune the lattice parameters and investigate the strain dependence of the Peierls transition in more details. 

In the present paper we investigate TTF-TCNQ thin films grown by physical vapor deposition on SrLaGaO$_4$(100), SrLaAlO$_4$(100), MgO(100), MgF$_2$(001), MgF$_2$(100), Si(100)/SiO$_2$(285~nm), $\alpha$-Al$_2$O$_3$(11$\bar{2}$0), and NaCl(100) substrates. We have done x-ray diffractometry to determine the relative alignment of the TTF-TCNQ crystallographic planes with respect to the substrate surface. The morphology of the TTF-TCNQ thin films were studied by scanning electron microscopy. From electrical conductivity measurements we determined the Peierls transition temperature of the thin films under the influence of several critical factors. In particular, we demonstrate that the TTF-TCNQ film thickness and the substrate material do not cause significant changes of the Peierls transition temperature, while the defect density in the thin films introduced during the evaporation process is a factor, which leads to a noticeable change in the Peierls transition temperature.

This paper is organized as follow. In section~\ref{ssec:exper} we give a short overview of the experimental methods used in the present study. The results on the structural, morphological, and electrical conductivity analysis in TTF-TCNQ thin films are presented in section~\ref{ssec:Res}. The influence of critical factors in TTF-TCNQ thin films on the Peierls transition temperature is also discussed in section~\ref{ssec:Res}. In section~\ref{ssec:concl} we conclude and give an outlook for further studies.

\section{Experimental Methods}
\label{ssec:exper}

TTF-TCNQ thin films of various thicknesses (ranging from 175~nm to 3~$\mu$m) were prepared by physical vapor deposition from the as-supplied TTF-TCNQ powder (Fluka, purity $\geq$97.0$\%$) at a background pressure $\leq 3\times 10^{-7}$~mbar. The material was sublimated from a low-temperature effusion cell using a quartz liner at a cell temperatures of 90$\ldots$130~$^{\circ}$C.  The cell temperature was measured by a Ni-NiCr-thermocouple thermally coupled to the heated body of the effusion cell by copper wool.

{\it Ex-situ} cleaved NaCl(100), and as-supplied chemically cleaned MgF$_2$(001), MgF$_2$(100), SrLaGaO$_4$(100), SrLaAlO$_4$(100), MgO(100), Si(100)/SiO$_2$(285nm), and $\alpha$-Al$_2$O$_3$(11$\bar{2}$0) substrates were used in the experiments. SrLaGaO$_4$(100) and SrLaAlO$_4$(100) substrates were chosen because their lattice parameters match closely the \textit{a} and \textit{b} cell parameters of the TTF-TCNQ crystal \cite{SrLaGaO4,SrLaAlO4}.

For our study we fabricated two series of samples: the first series was used to investigate the influence of the substrate material and the thickness of TTF-TCNQ thin films on the Peierls transition temperature. Five stripes of TTF-TCNQ thin films with different thicknesses were deposited, corresponding to growth periods of 1 to 5 hours, respectively, at an effusion cell temperature of 110~$^{\circ}$C and substrate temperature of 26~$^{\circ}$C. The distance between the effusion cell orifice and the substrate in this case was 50~mm. The second series of samples was used to investigate the influence of the evaporation temperature, i.e. growth rate, on the temperature of the Peierls transition temperature and film morphology. For the second series of samples the TTF-TCNQ thin films were deposited on Si(100)/SiO$_2$(285~nm) substrates employing evaporation temperatures of T$_{evap}$=90~$^{\circ}$C, 110~$^{\circ}$C, 130~$^{\circ}$C, respectively.

X-ray diffractometry (XRD) was performed employing a Bruker D8 diffractometer with a Cu anode in parallel beam mode using a Goebel mirror for studying the out-of-plane preferential orientation of the TTF-TCNQ films and for line profile analysis \cite{X-ray}. Scanning electron microscopy (SEM) was done with a FEI xT Nova NanoLab 600 at 5\,kV and a beam current of 98\,pA. The thickness of TTF-TCNQ thin films was determined from cross sections fabricated by a Ga focused ion beam (FIB) operating at 30~kV with a beam current of 0.5~nA. Energy dispersive x-ray spectroscopy (EDX) at 5~kV and 0.5~nA was employed to find the chemical composition of the films.

The electrical conductivity measurements were carried out in a $^4$He cryostat with a variable temperature insert allowing to cool down the sample from room temperature (300~K) to 4.2~K. The cooling/heating rate was 1~K per minute. We employed the four-probe technique at a fixed bias voltage of 0.1~V corresponding to an electric field in the thin films that did not exceed 0.2~V/cm.

\section{Results}
\label{ssec:Res}

\subsection{Preferential growth of TTF-TCNQ thin films} \label{ssec:growth}

\begin{figure}[htb]
\includegraphics[width=0.8\textwidth]{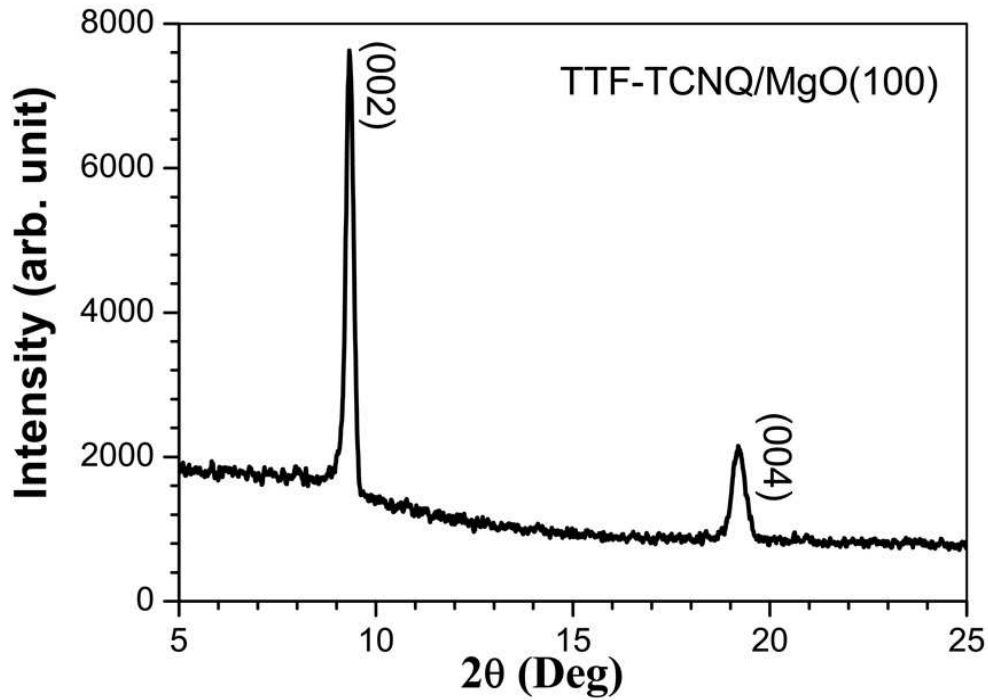}
\caption{X-ray diffraction pattern of TTF-TCNQ thin film deposited on MgO(100).}
\label{fig:X-ray_all}
\end{figure}

The XRD pattern recorded for a typical TTF-TCNQ thin film grown on MgO(100) is shown in Fig.~\ref{fig:X-ray_all}. Similar XRD patterns were obtained for TTF-TCNQ thin films grown on other substrates studied in this paper. Figure~\ref{fig:X-ray_all} shows that the reflections from the (00$\ell$) crystallographic plane of TTF-TCNQ dominate for even numbers of $\ell$, indicating that the $\textit{(ab)}$-molecular planes of the TTF-TCNQ thin film are aligned parallel to the substrate surface as illustrated in Fig.~\ref{fig:sturcture}. The observed preferential out-of-plane growth orientation for TTF-TCNQ thin films is in a good agreement with the results obtained in earlier studies on TTF-TCNQ thin films deposited on glass, sapphire, halide substrates etc. \cite{Chen_sapphire,TTF-TCNQ-halide,CVD_TTT-TCNQ,Fraxedas_OCVD,CVD_2009_TTF-TCNQ}.

\subsection{Morphology of TTF-TCNQ thin films} \label{ssec:morph}

\begin{figure}[htb]
\includegraphics[width=0.8\textwidth]{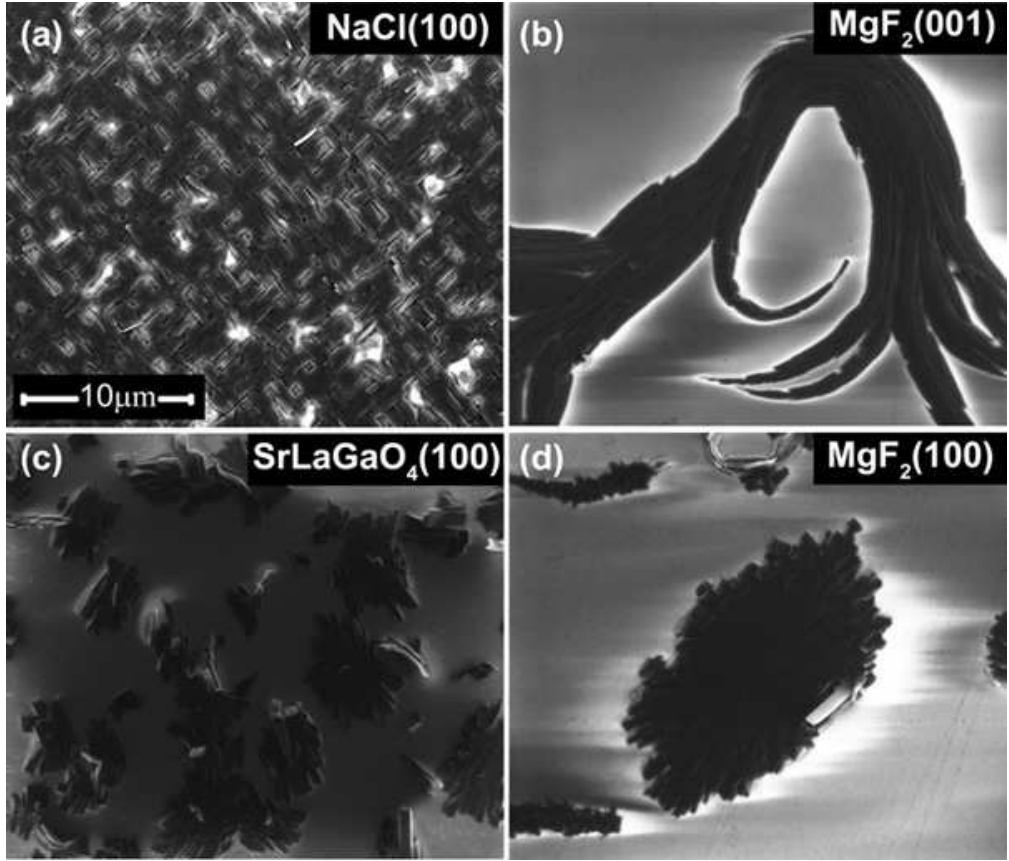}
\caption{SEM images of TTF-TCNQ thin films grown on (a) NaCl(100), (b) MgF$_2$(001), (c) SrLaGaO$_4$(100) and (d) MgF$_2$(100) substrates after 1~hour of deposition at 110~$^{\circ}$C effusion cell temperature.}
\label{fig:SEM_1h}
\end{figure}

\begin{figure}[htb]
\includegraphics[width=0.8\textwidth]{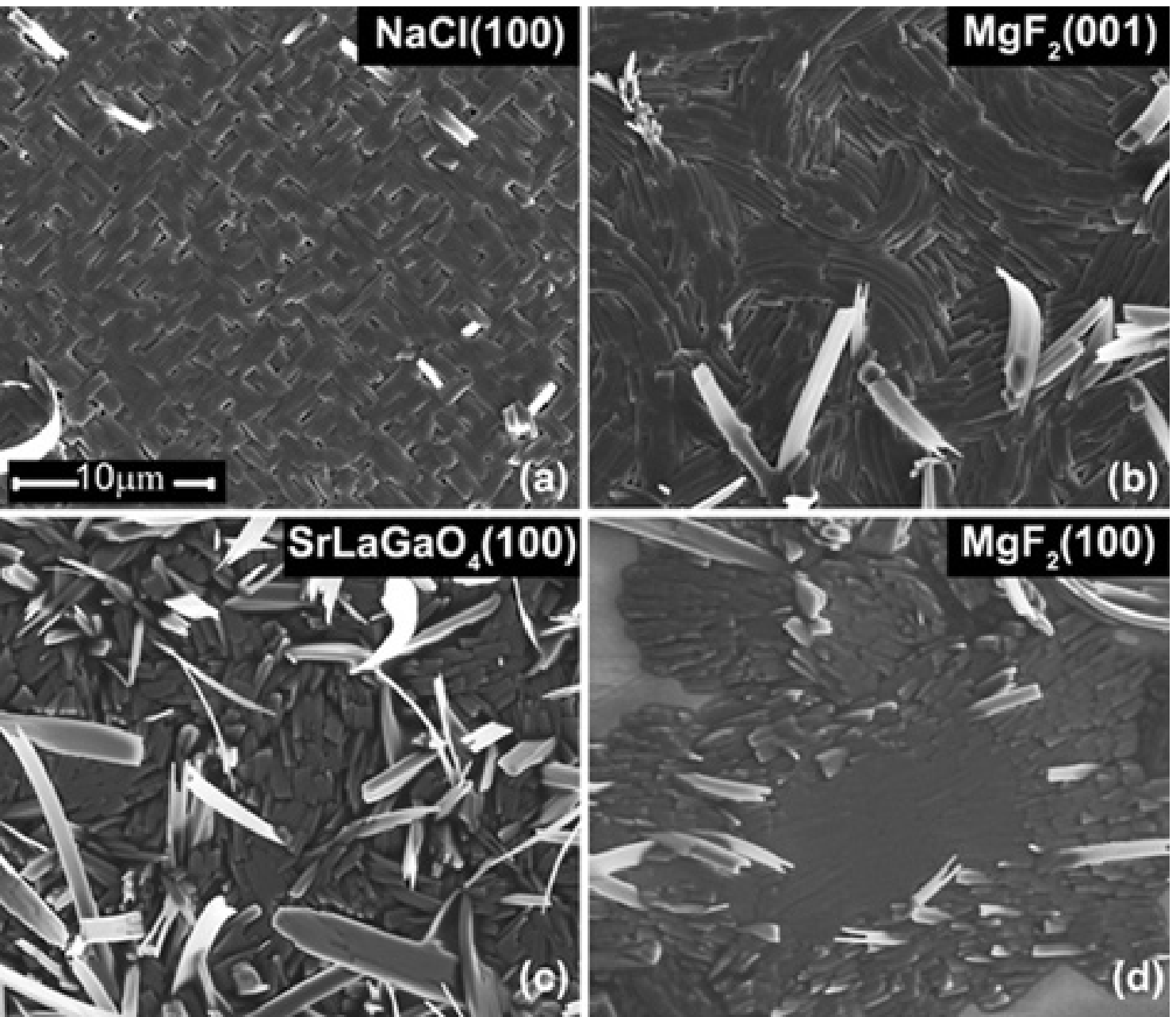}
\caption{SEM images of TTF-TCNQ thin films grown on (a) NaCl(100), (b) MgF$_2$(001), (c) SrLaGaO$_4$(100), and (d) MgF$_2$(100) substrates after 3~hour of deposition at 110~$^{\circ}$C effusion cell temperature.}
\label{fig:SEM_3h}
\end{figure}

\begin{figure}[htb]
\includegraphics[width=0.8\textwidth]{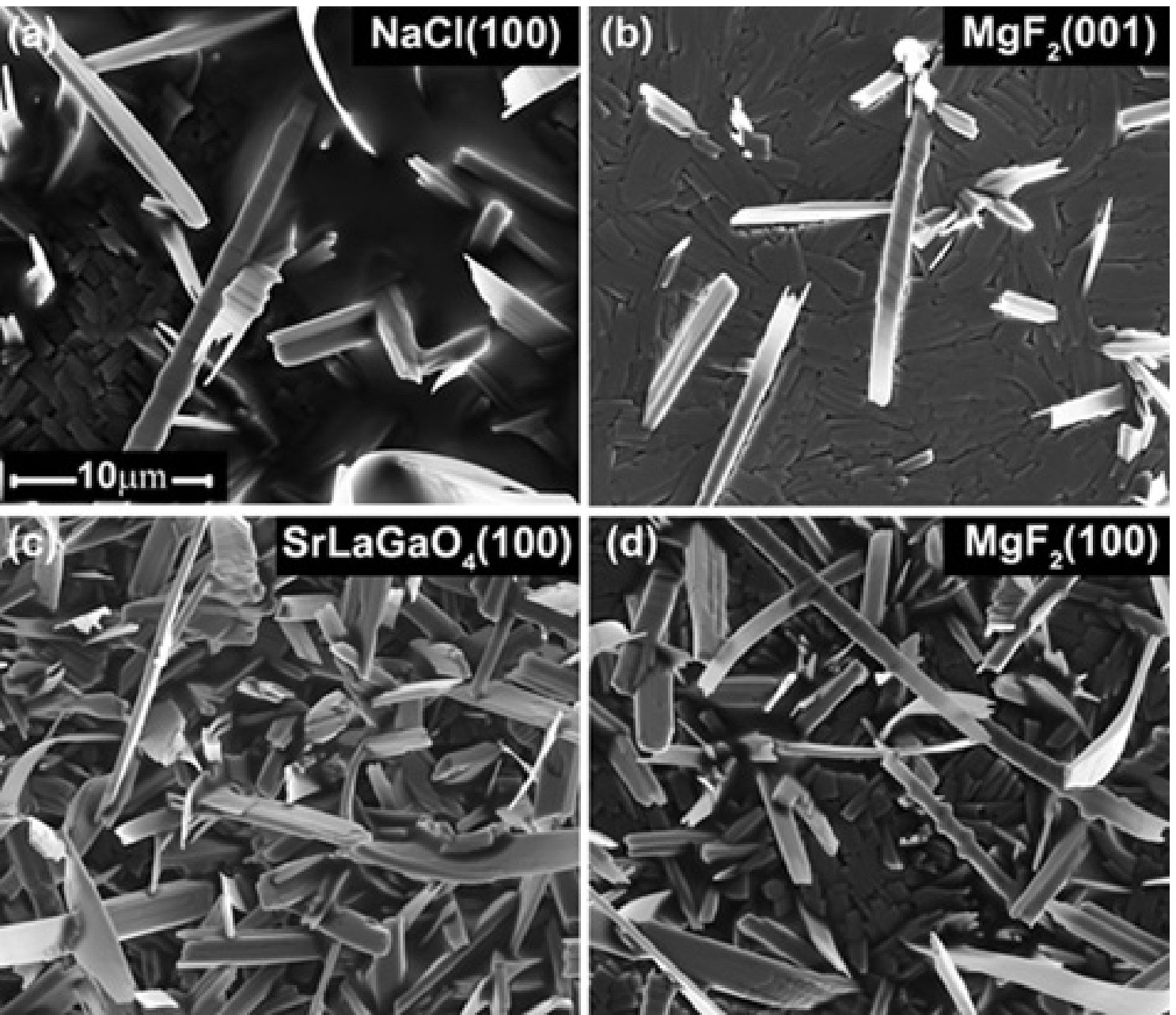}
\caption{SEM images of TTF-TCNQ thin films grown on (a) NaCl(100), (b) MgF$_2$(001), (c) SrLaGaO$_4$(100) and (d) MgF$_2$(100)) substrates after 5~hour of deposition at 110~$^{\circ}$C effusion cell temperature.}
\label{fig:SEM_5h}
\end{figure}

The morphology of the TTF-TCNQ thin films was investigated by SEM. Depending on the substrate material and orientation of the substrate, the thin films show several preferential growth modes. Figures~\ref{fig:SEM_1h}, \ref{fig:SEM_3h}, and \ref{fig:SEM_5h} present the morphology of the TTF-TCNQ thin films grown on NaCl(100), MgF$_2$(100), SrLaGaO$_4$(100) and MgF$_2$(001) substrates after 1, 3, and 5 hours of TTF-TCNQ deposition, respectively. From Figs.~\ref{fig:SEM_1h}, \ref{fig:SEM_3h}, and \ref{fig:SEM_5h} one can distinguish three growth modes: (1) thin films grown on the NaCl(100) substrate have two preferential orientations of the TTF-TCNQ crystals; (2) thin films grown on MgF$_2$(001) do not have any preferential orientation of the crystals, although the crystals selfassemble flat on the substrate surface; (3) islands without any preferential orientation of TTF-TCNQ crystals grow on the MgF$_2$(100) and on the SrLaGaO$_4$(100) substrates.

TTF-TCNQ thin films of a certain thickness selfassemble on the NaCl(100) substrate into a characteristic two-domain pattern as seen in Figs.~\ref{fig:SEM_1h}a and \ref{fig:SEM_3h}a. This result is in agreement with earlier investigations \cite{TTF-TCNQ-halide,Cond-morph-Volman,Indranil}. After the film thickness reaches a critical value, the thin film morphology acquires a pronounced 3D character (see Fig.~\ref{fig:SEM_5h}a). In this example, the critical thickness of the TTF-TCNQ thin film is about 600~nm.

\begin{figure}[htb]
\includegraphics[width=0.8\textwidth]{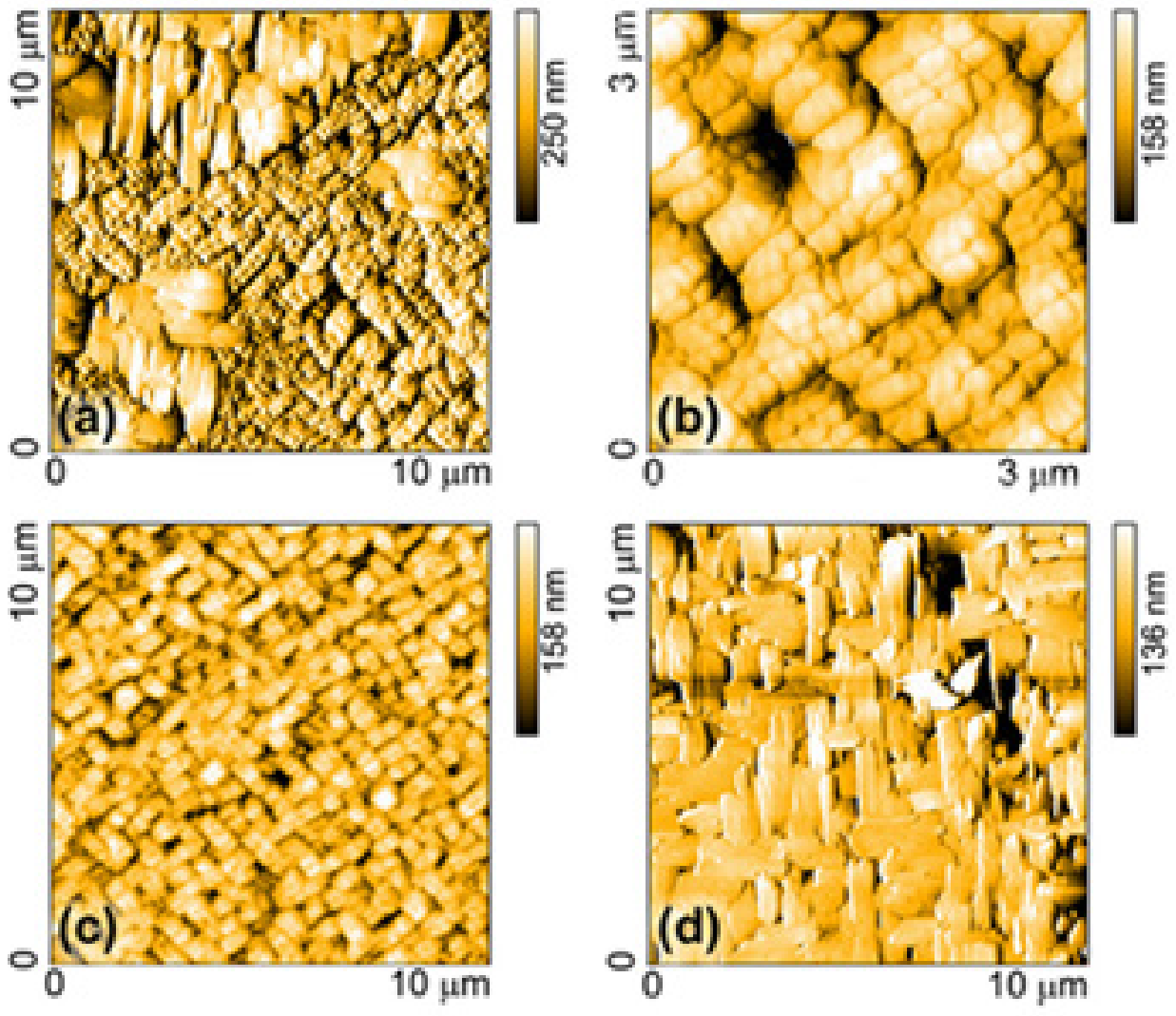}
\caption{AFM images of the TTF-TCNQ thin film grown on a NaCl(100) substrate after 1~hour of deposition at 110~$^{\circ}$C effusion cell temperature. Plot (a) illustrates the mixed area of the TTF-TCNQ thin film, where both, TCNQ and TTF-TCNQ crystallites are observed. Plots (b) and (c) illustrate regions of the lower layer of the thin film, which to a large extent consists of TCNQ crystals. Plot (d) illustrates selected regions showing the upper layer of the thin film, consisting of TTF-TCNQ crystallites.}
\label{fig:AFM_NaCl}
\end{figure}

The thickness of the TTF-TCNQ film grown during 1~hour deposition on NaCl(100) is $\sim$175~nm as determined by FIB cross sectioning. In this case AFM measurements allowed us to distinguish two different areas in the film's morphology (see Fig.~\ref{fig:AFM_NaCl}), which are hardly detectable in SEM measurements due to the charging effect. The first area corresponds to the top layer of the thin film where TTF-TCNQ crystallites with their b-axis oriented parallel to the $\langle110\rangle$ and $\langle\bar{1}10\rangle$ axes of the NaCl(100) substrate could be identified (see Fig.~\ref{fig:AFM_NaCl}d). The composition TTF:TCNQ $\approx$1:1 was verified by selected area EDX on these crystallites. 
The second area observed in AFM corresponds to a layer below the TTF-TCNQ crystallites of the grown thin film. The orientation of the crystallites in this region is parallel to the $\langle010\rangle$ and $\langle001\rangle$ directions of the NaCl(100) substrate (see Fig.~\ref{fig:AFM_NaCl}b-c). The chemical composition of the crystallites in the lower layer of the thin film was measured by EDX and showed that this layer consists to a large extent of TCNQ molecules. This result is in agreement with \cite{TCNQ_first_layer}, where it was demonstrated that during the growth process of TTF-TCNQ thin films on KaCl(100) a wetting layer of TCNQ is formed on the substrate surface.

In contrast to our observation on the NaCl(100) substrate, TTF-TCNQ forms flat arrays of crystallites without a preferred orientation on the MgF$_2$(001) substrate (see Fig.~\ref{fig:SEM_1h}b, \ref{fig:SEM_3h}b, and \ref{fig:SEM_5h}b ). This growth behavior is similar to TTF-TCNQ thin film self organization on KCl(100) kept at an elevated temperature of 325~K and annealed at 360~K \cite{TTF-TCNQ-halide}. Therefore, an oriented growth of the TTF-TCNQ thin film may be expected to occur if grown on a MgF$_2$(001) substrate kept at reduced temperature. However, the self organization process of the TTF-TCNQ thin films is apparently much more complicated than expected, and the cooling of the MgF$_2$(001) substrate down to 260~K during the evaporation did not lead to the formation of an oriented film.

\begin{figure}[htb]
\includegraphics[width=1\textwidth]{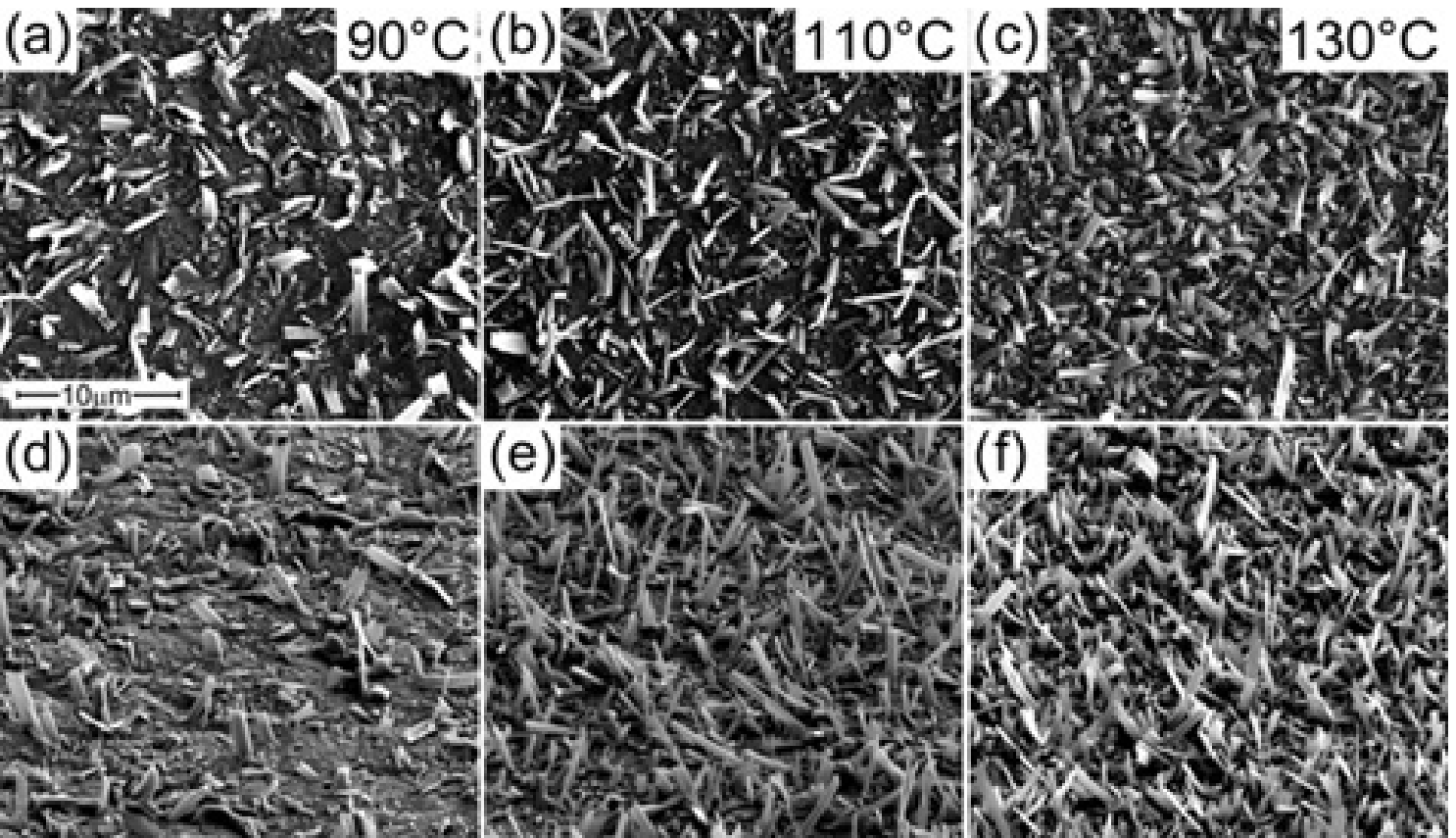}
\caption{SEM images of TTF-TCNQ thin films with identical thicknesses grown on Si(100)/SiO$_2$ recorded for samples aligned perpendicular (a-c) and with 52~$^{\circ}$ tilt (d-f) with respect to the electron beam for different evaporation temperatures: (a) and (d) for 90~$^{\circ}$C; (b) and (e) for 110~$^{\circ}$C; (c) and (f) for 130~$^{\circ}$C.}
\label{fig:SI_SiO2}
\end{figure}

The morphology of TTF-TCNQ thin films grown on the SrLaGaO$_4$(100) and MgF$_2$(100) substrates is shown in Figs.~\ref{fig:SEM_1h}c-d, \ref{fig:SEM_3h}c-d and \ref{fig:SEM_5h}c-d. A similar morphology of the TTF-TCNQ thin film is also typical for thin films grown on SrLaAlO$_4$(100), MgO(100), $\alpha$-Al$_2$O$_3$(11$\bar{2}$0) and Si(100)/SiO$_2$(285~nm) substrates, studied in this work. For the above mentioned substrates, the growth of the thin film originates from TTF-TCNQ islands (see Fig.~\ref{fig:SEM_1h}c-d). With increasing deposition time the islands coalesce and pronounced three-dimensional, bar-shaped TTF-TCNQ crystallites are formed (see Fig.~\ref{fig:SEM_3h}c-d and Fig.~\ref{fig:SEM_5h}c-d). Although, the discussed substrates lead to the formation of thin films with similar morphology, the sticking coefficient of the TTF-TCNQ molecules on the different substrates varies, thereby influencing the effective thin film growth rate. This was not studied in more detail.

An investigation of the influence of the evaporation temperature on the TTF-TCNQ thin film morphology was performed for the Si(100)/SiO$_2$(285~nm) substrate. Figure~\ref{fig:SI_SiO2} shows a series of SEM images recorded for TTF-TCNQ thin films grown on Si/SiO$_2$ at evaporation temperatures of 90~$^{\circ}$C, 110~$^{\circ}$C, and 130~$^{\circ}$C. 
From inspection of Fig.~\ref{fig:SI_SiO2} follows that the size of the TTF-TCNQ crystallites varies with the evaporation temperature, such that the higher the evaporation temperature the higher the crystallite density.

\subsection{The influence of the substrate material and film thickness on electrical conductivity of TTF-TCNQ thin films}
\label{ssec:conductivity}

Already from electrical conductivity measurements at room temperature it is apparent that the growth mode of the TTF-TCNQ thin films exerts a strong influence on their electrical properties. The minimal (${\sigma}_{min}$) and the maximal (${\sigma}_{max}$) values of the room temperature conductivity measured for various TTF-TCNQ thin films are given in Tab.~\ref{tab:cond}. As discussed in Sec.~\ref{ssec:morph}, the substrate influences the film morphology, which in turn determines the electrical conductivity. Consequently, the conductivity of the TTF-TCNQ thin films is substrate dependent, as seen from Tab.~\ref{tab:cond}. The maximal conductivity is observed for TTF-TCNQ thin films grown on the NaCl(100) and MgF$_2$(001) substrates. This is mainly due to the better quality of the TTF-TCNQ films having enhanced ordering of the crystallites. Nevertheless, the conductivity of the studied TTF-TCNQ thin films ($\sigma$) is in either case significantly smaller than the conductivity of the TTF-TCNQ single crystal ($\sigma_b$) and has a thermo-activated behavior presumably due to the high density of grain boundaries.

\begin{table}[htb]
\caption{The minimal (${\sigma}_{min}$) and the maximal (${\sigma}_{max}$) values of the room temperature conductivity ${\sigma}$ recorded for TTF-TCNQ thin films grown on different substrates. 5 different samples were measured for each substrate. The reference  conductivity of TTF-TNCQ for single crystals was measured as $\sigma_b\approx$500~($\Omega$~cm)$^{-1}$, $\sigma_a\approx$3~($\Omega$~cm)$^{-1}$~\cite{conductivity_ratio}.}
\label{tab:cond}
\begin{center}
\begin{ruledtabular}
\begin{tabular}{lll}
Substrate & ${\sigma}_{min}$, ($\Omega$ cm)$^{-1}$& ${\sigma}_{max}$, ($\Omega$ cm)$^{-1}$ \\
\hline
NaCl(100)&1.32&30
\\
\\SrLaGaO$_4$(100) & 0.08 �&0.35
\\
\\SrLaAlO$_4$(100)& �0.02& 0.03
\\
\\ MgO(100)  &0.48 & 1
\\
\\MgF$_2$(001)�& 3.19& 9.68
\\
\\MgF$_2$(100)& 0.58& 0.86�
\\
\\$\alpha$-Al$_2$O$_3$(11$\bar{2}$0)&0.01&2.4

\end{tabular}
\end{ruledtabular}
\end{center}
\end{table}

\begin{figure}[htb]
\includegraphics[width=1\textwidth]{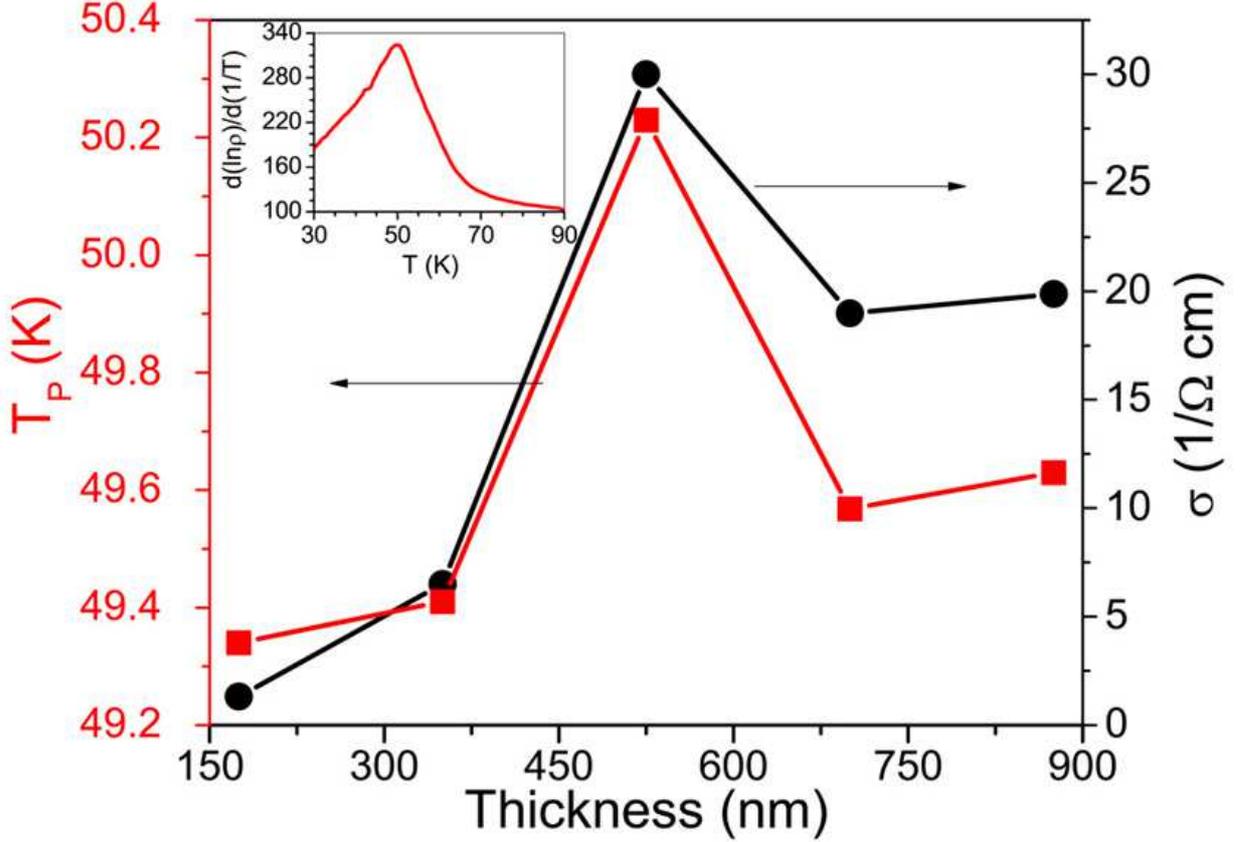}
\caption{Electrical conductivity (circles) and the Peierls transition temperature (squares) measured for TTF-TCNQ thin films with different thicknesses grown on the NaCl(100) substrate. The inset shows the dependence of resistivity derivative, Eq.~(\ref{eq:func}), on temperature extracted for the TTF-TCNQ thin film formed on the NaCl(100) substrate after 4 hours of the deposition.}
\label{fig:cond_thickn}
\end{figure}

As follows from the performed analysis, the TTF-TCNQ thin films grown on the NaCl(100) substrate form planar two-domain structures. Therefore, we studied the influence of the film thickness on the room temperature conductivity for this substrate. We found that the room temperature conductivity depends non-monotonously on the film thickness, as is presented in Fig.~\ref{fig:cond_thickn}. The room temperature conductivity shows a maximal value at a thickness of about 525~nm. The likely reason for this behavior can be deduced from AFM and SEM measurements (see Figs.~\ref{fig:SEM_1h}-\ref{fig:AFM_NaCl}). The increase of the thin film thickness eventually leads to the coalescence of TTF-TCNQ islands on the substrate surface, enhancing the conductivity. When this coalescence occurs the film thickness is about 500~nm and the electrical conductivity is maximal. A further increase of the film thickness results in the decrease of the conductivity because the growth mode is changed into a disordered growth that yields three dimensional bar-shaped crystallites partly pointing out of the surface. 
A similar dependence was reported in \cite{Cond-morph-Volman} for TTF-TCNQ thin films grown on alkali halide and glass substrates. Although the non-monotonic dependence of the conductivity of TTF-TCNQ films on thickness was also observed for all other substrate investigated in this work, a reliable thickness measurements is severely hindered or even impossible due to the irregular and very pronounced three-dimensional growth already in early stages. This does not allow us to deduce a reliable conductivity-thickness dependence as was done for the NaCl(100) substrate.

\subsection{Dependence of the Peierls transition temperature on the defect density in TTF-TCNQ thin films}
\label{ssec:defects}
The Peierls transition temperature $T_P$ can be extracted from the temperature dependence of the electrical conductivity $\sigma(T)$ (see inset to Fig.~\ref{fig:cond_thickn}). It corresponds to the temperature at which the function

\begin{equation}
\label{eq:func}
\frac{\mathrm{d}(\ln\rho(T))}{\mathrm{d}(1/T)}=\frac{\mathrm{d}(\ln1/\sigma(T))}{\mathrm{d}(1/T)},
\end{equation}
\noindent
exhibits a maximum \cite{Gruner}. Here $\rho(T)$ is the resistivity of the thin film.

We find that the Peierls transition temperatures for all TTF-TCNQ thin films grown at the evaporation temperature of  110$^{\circ}$C is close to 50~K. Note that no systematic dependence of the Peierls transition temperature on the substrate material and thickness could be observed. Films grown on NaCl(100) substrate are an exception from this rule and will be discussed separately. This observation can be qualitatively explained by the relatively weak forces acting between the substrate and the TTF-TCNQ thin films. Indeed, exempting the NaCl(100) substrate, the forces acting between the TTF-TCNQ thin film and the substrate are mainly van der Waals in nature, being significantly weaker than the intermolecular forces in the thin film and leading to a three dimensional Volmer-Weber growth mode \cite{nucleation_and_growth}. The van der Waals forces are sufficient for the formation of TTF-TCNQ thin film on a substrate surface, but these forces are not strong enough to clamp the thin film to the substrate. Therefore, the cooling of the sample does not induce tensile biaxial strain in the thin film, and no significant influence of the substrate material on the Peierls transition temperature  is expected.

The binding between the NaCl(100) substrate and the TTF-TCNQ thin film has an ionic component. Therefore, an influence of the thin film thickness on the Peierls transition temperature is likely to occur. With increase of the thin film thickness the interaction of the growing layers with the substrate weakens.  The thicker the film the more dislocations nucleate to compensate for the lattice misfit. At some critical thickness the gained energy is enough to cross the Peierls barrier and the lattice constants of the film relax to their bulk value (see e.g. \cite{Huth-Peierls-barrier} and references therein). This critical thickness for TTF-TCNQ thin films in the experiment is presumably about 525~nm including the TCNQ wetting layer. The dependence of the Peierls transition temperature on TTF-TCNQ film thickness grown on NaCl(100) is shown in Fig.~\ref{fig:cond_thickn}. In the insert the derivative of the resistivity on temperature for one chosen thin film is presented for illustrative purposes. As is evident, the dependence of the conductivity on the film thickness is followed by the dependence of the Peierls transition temperature on thickness. We suggest that this observation can be explained by a correlation between the defect density and the Peierls transition temperature as follows. As known for TTF-TCNQ single crystals \cite{Chiang} an increase of the defect density leads to a decrease of the Peierls transition temperature. Here, the interaction of the thin film with the substrate can induce defects in the film. The inverse conductivity can be used as a measure of the defect density \cite{defect_measure}. Figure~\ref{fig:cond_thickn} shows that the Peierls transition temperature increases with increasing electrical conductivity (or, equivalently, decreasing defect density) and this is in agreement with the result observed for TTF-TCNQ single crystals  \cite{Chiang}.

One indicator of the defect presence is the formation of defect-induced microstrain in the films which can be deduced from a line profile analysis \cite{X-ray} of the x-ray diffraction patterns for films grown on NaCl(100). The analysis provides $K_D \varepsilon_{rms}$, where $\varepsilon_{rms}$ is the root mean square microstrain in the films for different film thicknesses and $K_D$  is a scaling factor depending on the nature of the microstructural changes \cite{X-ray}. $\varepsilon_{rms}$ is defined as

\begin{equation}
 \varepsilon_{rms}=\sqrt{\langle\varepsilon^2\rangle},
\end{equation}
\noindent where $\varepsilon=\bigtriangleup d/d_0$, $\bigtriangleup d$ is the variation of interplanar spacing and $d_0$ is the undistorted spacing.

\begin{figure}[tb]
\includegraphics[width=1\textwidth]{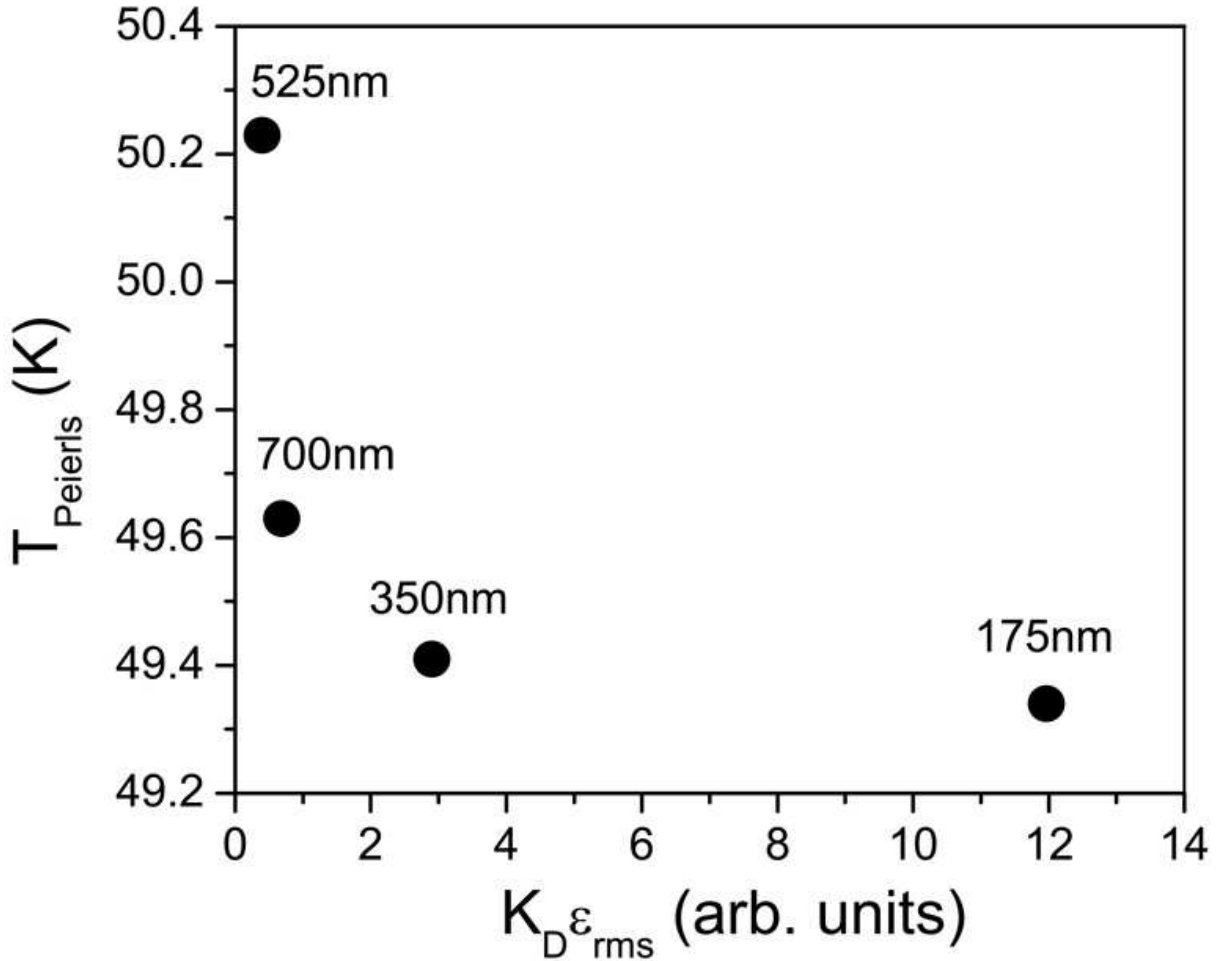}
\caption{The dependence of the Peierls transition temperature on room mean square microstrain in TTF-TCNQ thin films, extracted from line profile analysis. K$_D$ is a scaling factor depending on the nature of the microstructural changes \cite{X-ray}. The thin film thickness is indicated.}
\label{fig:microstrain}
\end{figure}

The result of the analysis is shown in Fig.~\ref{fig:microstrain}. 
The microstrain is usually caused by microstructural changes in the sample structure, i.e. non-uniformity of crystallite shape, dislocations etc. \cite{X-ray}. As follows from Fig.~\ref{fig:microstrain}, thinner films experience a larger microstrain and, therefore, possesses more defects presumably because the influence of the substrate is increasing as the thickness of the thin film decreases. The Peierls transition temperature also increases with the strain decrease (see Fig.~\ref{fig:microstrain}) for thin films with a thickness up to 525~nm, supporting the hypothesis that the defects have a critical impact on this transition in thin films. The increase of the microstrain in the case of the TTF-TCNQ film with thickness of 700~nm as compared to the film with thickness of 525~nm is caused by the interplay of two effects: a decrease of the microstrain due to the relaxation of the film by dislocation formation and an increase of strain caused by the increasingly disordered film microstructure.

\begin{figure}[htb]
\includegraphics[width=1\textwidth]{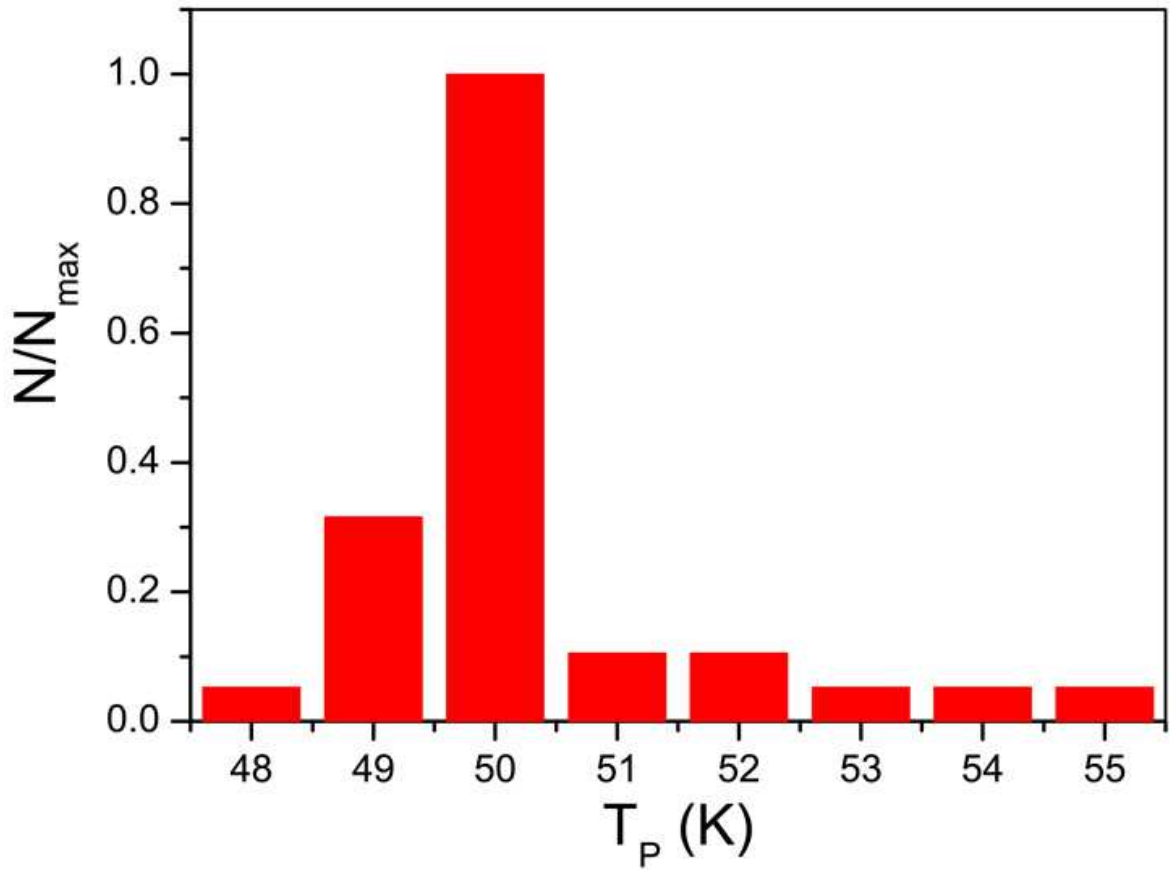}
\caption{Normalized temperature distribution of the Peierls transition temperature of  different TTF-TCNQ thin films grown at different conditions. $N_{max}$ is a number of the samples for which the distribution has a maximum. The total number of investigated samples is 40.}
\label{fig:T-P-distribution}
\end{figure}

Figure~\ref{fig:T-P-distribution} shows the normalized temperature distribution of the Peierls transition temperature. The Peierls transition temperature exhibits a pronounced maximum at $\sim$ 50~K which is smaller than the one reported for single crystals $\approx$54~K \cite{TTF-TCNQ-first,TTF-TCNQ_giant}. 

It is worth noting that if one considers thin films of the blue-bronze, which also exhibit charge density waves, the Peierls transition temperature for this material is also shifted towards lower values as compared to single crystals \cite{blue-bronze}. Therefore, we suggest that the shift observed here can be ascribed to the presence of defects in the TTF-TCNQ thin films.

To estimate the defect concentration from the Peierls transition temperature shift we use the microscopic calculation approach introduced in \cite{Abrikosov,Patton}. The decrease of the Peierls transition temperature due to defects obeys the following relation:

\begin{equation}
\label{eq:T-per}
 \ln\left(\dfrac{T_{P0}}{T_P}\right)=\Psi\left(\dfrac{1}{2}+\dfrac{\hbar}{2\pi k_B \tau T_P}\right)-\Psi\left(\dfrac{1}{2}\right),
\end{equation}

\noindent where $T_{P0}$ and $T_P$ are the transition temperatures in the absence of defects and with defects, respectively. $k_B$ is the Boltzmann constant, $\hbar$ is the reduced Planck constant. $\Psi(x)$ is the digamma function, $\tau$ is the scattering time due to the presence of defects and can be estimated as

\begin{equation}
\label{eq:mean}
l=\tau v_F.
\end{equation}
\noindent Here $l$ is the average distance between the defects and $v_F$ is the Fermi velocity. From Eq.~(\ref{eq:T-per}) for $T_{P0}$=54~K and $T_P$=50~K one obtains $\tau=1.52\times 10^{-12}$~s. The Fermi velocity for TTF-TCNQ is $\sim 1.82\times 10^7$~cm/s \cite{Pouget}. Using Eq.~(\ref{eq:mean}) we obtain $l\approx2.7\times 10^{-5}$~cm which corresponds to 0.14~\% defect concentration. This can be compared with the single crystal analysis performed in \cite{Chiang} for different defect concentrations induced by deuteron irradiation of the sample: a decrease of the Peierls transition temperature of 4~K, as observed in our experiments, corresponds to a defects concentration of about 0.1$\ldots$1\%. This agrees favorably with our analysis presented here. The uncertainty in defect concentration for single crystal stems from the uncertainty in converting the deuteron flux into induced defect concentration, as detailed in \cite{Chiang}.

\section{Conclusion}
\label{ssec:concl}

In this paper we have investigated the preferential growth directions of TTF-TCNQ thin films for various substrate materials and crystallographic orientations. The morphology of the thin films was studied for different film thicknesses. It was shown that for some critical thickness the morphology of the TTF-TCNQ films changes from two-dimensional into a three-dimensional structure only for grown on NaCl(100) and MgF$_2$(001).

The Peierls transition temperature was analyzed for TTF-TCNQ thin films of varies thicknesses, grown on different substrate materials and orientations. We demonstrated that the influence of most substrate materials on the Peierls transition temperature is negligible and no clamping between thin film and substrate takes place. Contrary, if the interaction between the TTF-TCNQ thin film and the substrate is strong, as is the case for growth on NaCl(100) substrates, the Peierls transition temperature depends on the thin film thickness. The interaction between thin film and substrate causes the occurrence of microstrain in thin films.

We furthermore demonstrated that the Peierls transition temperature of TTF-TCNQ thin films is shifted towards lower values, as compared to the TTF-TCNQ single crystals. The defects which emerge in TTF-TCNQ microcrystals due to the non-equilibrium growth process may destroy the long range order and be partly responsible for this temperature shift. Using a theoretical framework developed in  \cite{Abrikosov,Patton} we estimated the concentration of defects as $\sim$0.14\%.

This paper point towards some open questions. For example, it would be interesting to investigate the behavior of the Peierls transition in the TTF-TCNQ microcrystals and nanowires \cite{Ilia-PRB}, where the substrate can induce a stronger influence on the microcrystal, as compared to thin films, leading to uni- and biaxial strains in the system, as was done for $\kappa$-(BEDT-TTF)$_2$Cu$[$N(CN$)_2]$Br in \cite{Hiroshi_strain}. Such a study would have to be complemented by a theoretical analysis of the dependence of the Peierls transition temperature on uni- and bi-axial strain.

\section{Acknowledgment}

The authors are grateful to the Sonderforschungsbereich/Transregio 49 project for financial support of this work. They also thank Roland Sachser and Anastasia Cmyrev for their support in doing the conductivity and SEM measurements, Oleksandr Foyevtsov for his assistance in performing line profile analysis and Dr. Ilia Solov'yov for fruitful discussions and critical reading of the manuscript.

\end{document}